\documentstyle[nato,epsf,numreferences]{crckapb}   

\newcommand\al{\alpha}
\newcommand\ga{\gamma}
\newcommand\de{\delta}
\newcommand\ep{\epsilon}
\newcommand\ze{\zeta}
\newcommand\et{\eta}
\newcommand\th{\theta}
\newcommand\vt{\vartheta}

\newcommand\ka{\kappa}
\newcommand\la{\lambda}
\newcommand\rh{\rho}
\newcommand\si{\sigma}
\newcommand\ta{\tau}

\newcommand\ph{\phi}
\newcommand\vp{\varphi}
\newcommand\ch{\chi}
\newcommand\ps{\psi}

\newcommand\De{\Delta}

\newcommand\Ph{\Phi}

\newcommand\Om{\Omega}

\newcommand\pa{\partial}
\newcommand\na{\nabla}
\newcommand\bna{\nabla}
\newcommand\hb{\hbar}

\newcommand\ie{{\em i.e.}}
\newcommand\eg{{\em e.g.}}

\newcommand\be{\begin{equation}}
\newcommand\ee{\end{equation}}
\newcommand\ba{\begin{eqnarray}}
\newcommand\ea{\end{eqnarray}}
\newcommand\pmat{\pmatrix}

\newcommand\X{\times}
\newcommand\ov{\over}

\newcommand\rms[1]{_{\rm #1}}
\newcommand\Tc{{T\rms{c}}}
\newcommand\kB{{k\rms{B}}}

\newcommand\tr{{\rm tr}}
\newcommand\gap{\;\lower3pt\hbox{$\buildrel > \over
{\scriptstyle\sim}$}\;}
\newcommand\lap{\;\lower3pt\hbox{$\buildrel < \over 
{\scriptstyle\sim}$}\;}
\newcommand\dg{^{\dagger}}
\newcommand\case[2]{\textstyle{#1\over#2}}
\newcommand\half{\case{1}{2}}
\newcommand\ap{\approx}

\newcommand\<{\langle}
\renewcommand\>{\rangle}
\newcommand\cd{\cdot}
\newcommand\ua{\uparrow}
\newcommand\da{\downarrow}

\newcommand\PL{{\em Phys.\ Lett.\ }}
\newcommand\PRD{{\em Phys.\ Rev.\ D\ }}
\newcommand\PRL{{\em Phys.\ Rev.\ Lett.\ }}
\newcommand\NP{{\em Nuc.\ Phys.\ }}

\newcommand\MPLA{{\em Mod.\ Phys.\ Lett.\ A\ }}

\newcommand\Nat{{\em Nature\ }}

\newcommand\Hh{\hat{H}}
\newcommand\Ls{{L\rms{str}}}
\newcommand\Nh{\hat{N}}
\newcommand\TG{{T\rms{G}}}
\newcommand\Uh{\hat{U}}
\newcommand\ah{\hat{a}}
\newcommand\bA{{\bf A}}
\newcommand\bd{{\bf d}}
\newcommand\bh{{\bf h}}
\newcommand\bhat{\hat{b}}
\newcommand\bk{{\bf k}}
\newcommand\bl{{\bf l}}
\newcommand\bm{{\bf m}}
\newcommand\bn{{\bf n}}

\newcommand\br{{\bf r}}
\newcommand\bv{{\bf v}}
\newcommand\bvn{{\bf v}\rms{n}}
\newcommand\bvs{{\bf v}\rms{s}}
\newcommand\bD{{\bf D}}
\newcommand\bH{{\bf H}}
\newcommand\bI{{\bf I}}
\newcommand\bJ{{\bf J}}
\newcommand\bM{{\bf M}}
\newcommand\bMh{\hat{\bf M}}
\newcommand\bL{{\bf L}}
\newcommand\bQ{{\bf Q}}
\newcommand\bR{{\bf R}}
\newcommand\bS{{\bf S}}
\newcommand\bSh{\hat{\bf S}}

\newcommand\bZ{{\bf Z}}
\newcommand\bO{{\bf 0}}
\newcommand\bone{{\bf 1}}
\newcommand\cs{c\rms{s}}
\newcommand\eZ{{\ep\rms{Z}}}
\newcommand\kF{k\rms{F}}
\newcommand\ms{{m\rms{s}}}
\newcommand\mv{{m\rms{v}}}
\newcommand\ns{{n\rms{s}}}

\newcommand\phh{\hat{\ph}}
\newcommand\tq{{\ta\rms{q}}}
\newcommand\tZ{{t\rms{Z}}}
\newcommand\vc{{v\rms{c}}}
\newcommand\vcn{{v\rms{cn}}}
\newcommand\vs{{v\rms{s}}}
\newcommand\xe{\xi\rms{eq}}
\newcommand\xs{\xi\rms{str}}
\newcommand\xp{\xi_\ph}
\newcommand\xZ{\xi\rms{Z}}

\begin{opening}

\title{SYMMETRY BREAKING AND DEFECTS}

\author{T.W.B. KIBBLE}
\institute{Blackett Laboratory, Imperial College,\\
London SW7 2BW, United Kingdom}

\end{opening}

\runningtitle{SYMMETRY BREAKING AND DEFECTS}

\begin{document}

\section{Introduction}

Symmetry-breaking phase transitions are ubiquitous in
condensed matter systems and in quantum field theories. 
There is also good reason to believe that they feature
in the very early history of the Universe.  At many such
transitions topological defects of one kind or another
are formed.  Because of their inherent stability, they
can have important effects on the subsequent behaviour
of the system. 

In the first of these lectures I shall review a number
of examples of spontaneous symmetry breaking, many of
which will be discussed in more detail by other
lecturers, and discuss their general features.  The
second lecture will be mainly devoted to the conditions
under which topological defects can appear and their
classification in terms of homotopy groups of the
underlying vacuum manifold.  In my final lecture, I will
discuss the `cosmology in the laboratory' experiments
which have been done to try to test some of the ideas
thrown up by discussions of defect formation in the
early Universe by looking at analogous processes in
condensed-matter systems.

\section{Spontaneous symmetry breaking}

Often a system has symmetries that are not shared by its
ground state or vacuum state.  This is the phenomenon of
{\em spontaneous symmetry breaking}.  It always implies
a degeneracy of the ground state.  In this lecture, I
want to discuss a number of simple examples of this
phenomenon. 

The key signature of spontaneous symmetry breaking is
the existence of some operator $\phh$, the {\em order
parameter} (for example the magnetization $\bMh$ in a
ferromagnet) whose ground-state expectation value is not
invariant.  Typically, the equilibrium state at high
temperature is invariant --- for example, above the
Curie temperature $\Tc$ the expectation value of $\bMh$
vanishes.  Symmetry breaking appears as the system is
cooled below $\Tc$.  It spontaneously acquires
magnetization in some random direction.

\subsection{Ferromagnet}

As a first example, let us think about a Heisenberg
ferromagnet \cite{Yeo92}, a system of spins $\bSh_\br$ at
lattice sites $\br$, with Hamiltonian
 \be
\Hh = - {1\ov2}\sum_\br \sum_{\br'} J(|\br-\br'|)
\bSh_\br\cd\bSh_{\br'} - \bH\cd\sum_{\br}\bSh_\br,
 \label{fmHam}
 \ee
where $J$, assumed positive, is non-zero only for
neighbouring spins, and $\bH$ represents a possible
external magnetic field.  For non-zero $\bH$, the ground
state $|0\>$ of the system has all spins aligned in the
direction of $\bH$:
 \be
\<0|\bSh_\br|0\> = {1\ov2}\bh\cd\si, \qquad
\bh={\bH\ov|\bH|}.
 \ee   
If we take the infinite-volume limit and then let
$\bH\to\bO$, this remains true.

Clearly, with $\bH=\bO$, (\ref{fmHam}) is invariant under
all rotations. The symmetry group is $SO(3)$ or, rather,
its two-fold covering group $SU(2)$.  However,
$\<0|\bSh_\br|0\>$ is evidently not invariant.  Applying
the rotation operators to $|0\>$ yields an infinitely
degenerate set of ground states $|0_\bh\>$, labelled by
the directions of the unit vector $\bh$.  Note that the
scalar product of two distinct ground states tends to
zero in the infinite-volume limit:
 \be
|\<0_\bh|0_{\bh'}\>| = \cos^N{\th\ov2} \to 0 \quad {\rm
as} \quad N \to \infty,
 \ee
where $N$ is the number of lattice sites and $\th$ is the
angle between $\bh$ and $\bh'$.  In fact, these two
states $|0_\bh\>$ and $|0_{\bh'}\>$ belong to separate,
mutually orthogonal Hilbert spaces, with unitarily
inequivalent representations of the commutation
relations.  No operator constructed from a finite set
of spins has a nonvanishing matrix element between
$|0_\bh\>$ and $|0_{\bh'}\>$.  This is another
characteristic of spontaneously broken symmetry.

Note that physically, unless we introduce a
symmetry-breaking interaction with an external magnetic
field, the different ground states are
indistinguishable.  Indeed it is possible (though
not particularly helpful) to define an invariant ground
state, a uniform superposition of all $|0_\bh\>$.

\subsection{Free Bose gas}

Another example of spontaneous symmetry breaking is
Bose-Einstein condensation.  Let us first consider a
free Bose gas \cite{MarR02}, with Hamiltonian
 \be 
\Hh = \sum_{\bk} \ep_\bk \ah^{\dg}_\bk \ah_\bk,
 \ee
where $\ah_\bk$ and $\ah^{\dg}_\bk$ are the destruction
and creation operators for a particle of momentum
$\hb\bk$, satisfying the commutation relations
 \be 
[\ah_\bk, \ah^{\dg}_{\bk'}] = \de_{\bk,\bk'},
 \ee
and 
 \be
\ep_\bk = {\hb^2\bk^2\ov 2m}.
 \ee 
The relevant symmetry here is the phase symmetry
 \be
\ah_\bk\mapsto e^{i\vt}\ah_\bk, \qquad
\ah^{\dg}_\bk\mapsto e^{-i\vt}\ah^{\dg}_\bk,
 \label{Absymm}
 \ee
corresponding to the existence of a conserved particle
number,
 \be
\Nh=\sum_\bk \ah^{\dg}_\bk \ah_\bk.
 \ee

In the grand canonical ensemble, the mean occupation
number of mode $\bk$ is given by the Bose-Einstein
distribution
 \be
\<\ah^{\dg}_\bk \ah_\bk\> = N_\bk \equiv
{1\ov e^{\beta(\ep_\bk-\mu)}-1},
 \label{BEoccno}
 \ee
where $\beta=1/\kB T$ is the inverse temperature and
$\mu$ the chemical potential.  If the mean number density
$n=\<\Nh\>/\Om$ (where $\Om$ is the volume) is specified,
then the value of $\mu$ for any large temperature $T$ is
given by
 \be
n = {1\ov\Om}\sum_\bk {1\ov e^{\beta(\ep_\bk-\mu)}-1}
\ap \int {d^3\bk\ov(2\pi)^3} {1\ov
e^{\beta(\ep_\bk-\mu)}-1}.
 \label{BEn}
 \ee
As $T$ falls, $\mu$ increases.  Clearly, to avoid a
divergence, $\mu$ can never become positive, but
eventually it approaches zero at some critical
temperature $\Tc$.  At that point, the occupation number
(\ref{BEoccno}) diverges at $\bk=\bO$, so the integral
approximation in (\ref{BEn}) is no longer adequate.  We
have to separate the $\bk=\bO$ mode from the rest.  For
all the other values of $\bk$, we can still make the
integral approximation, and set $\mu=0$.  Thus we get
 \be
n = n_\bO + \int {d^3\bk\ov(2\pi)^3} {1\ov
e^{\beta\ep_\bk}-1}, \quad {\rm with}\quad
n_\bO \ap {1\ov-\Om\beta\mu}.
 \label{BEcond}
 \ee
For $T<\Tc$, $\mu$ becomes extremely small, of order
$-\kB T/n\Om$.  Thus, a macroscopically significant
fraction of all the particles is to be found in the
single zero-momentum mode. 

Note that from (\ref{BEcond}) we can calculate the
critical temperature $\Tc$ at which $n_\bO$ first
becomes large.  It is given by
 \be
n = {1\ov(2\pi)^2}\left(2m\kB \Tc\ov\hb^2\right)^{3/2}
\int_0^\infty {x^{1/2}dx\ov e^x-1},
 \ee
which yields
 \be
\Tc = {2\pi\ov\ze({3\ov2})^{2/3}} {\hb^2 n^{2/3}\ov m\kB}
= 3.31 {\hb^2 n^{2/3}\ov m\kB}.
 \label{BETc}
 \ee

\subsection{Degenerate ground state}

To see how a degenerate ground state emerges in this
case, it is more convenient to work in terms of the
scalar field 
 \be
\phh(\br) = {1\ov\Om^{1/2}}\sum_\bk \ah_\bk
e^{i\bk\cd\br}. 
 \ee
Then we can write
 \be
\Hh = \int_\Om d^3\br\;
{\hb^2\ov2m}\bna\phh\dg(\br)\cd\bna\phh(\br). 
 \ee
The phase symmetry (\ref{Absymm}) can now be expressed as
 \be
\phh(\br)\mapsto e^{i\vt}\phh(\br), \qquad
\phh\dg(\br)\mapsto e^{-i\vt}\phh\dg(\br).
 \label{phaserot}
 \ee

Now let us introduce an explicit symmetry-breaking source
term playing the same role as an external magnetic field
for the ferromagnet.  In this case, we take
 \be
\Hh_1 = -\int_\Om d^3\br [j^*\phh(\br)+j\phh\dg(\br)]
= -\Om^{1/2}(j^*\ah_\bO+j\ah\dg_\bO),
 \ee
where $j$ is a complex number, and consider the limit
$j\to0$ only {\em after} letting $\Om\to\infty$.  Here
$\Hh_1$ may be thought of as representing the
possibility of particle exchange with the environment. 
We then find that the density operator $\rh_\bO$ for the
zero-momentum mode is the same as with $j=0$ but with a
shifted field:
 \be
\rh_\bO = (1-e^{\beta\mu})e^{\beta\mu\bhat\dg_\bO\bhat_\bO},
 \ee
where
 \be 
\bhat_\bO = \ah_\bO + {\Om^{1/2}j\ov\mu}.
 \ee
To find $\mu$ we must again use (\ref{BEn}) and
(\ref{BEcond}) but appropriately modified, namely
 \ba
\Om n_\bO \!\!\!&=&\!\!\! \<\ah\dg_\bO\ah_\bO\>
= \<\bhat\dg_\bO\bhat_\bO\> + {\Om|j|^2\ov\mu^2}\nonumber\\
&=&\!\!\! {1\ov e^{-\beta\mu}-1} + {\Om|j|^2\ov\mu^2}.
 \ea
Clearly, as $\Om\to\infty$, the first term on the right
becomes negligible, and so
 \be
\mu\ap-{|j|\ov n_\bO^{1/2}}.
 \ee
(We assume for simplicity that $|j|\ll
\sqrt{n_\bO}/\beta$, so that $\mu$ remains small enough
to be negligible for the $\bk\ne\bO$ modes.)  Then we
find that the ground state for the zero-momentum mode is
a {\it coherent state} $|\ph\>$, such that
$\bhat_\bO|\ph\>=0$, in which the expectation value of
$\phh$ is non-zero, with a phase constrained by the
phase of $j$:
 \be
\<\phh(\br)\> = \ph \equiv \et e^{i\vt} =
n_\bO^{1/2}{j\ov|j|}.
 \ee

The coherent states may be labelled by the value of
$\ph$, and are given explicitly by
 \be
|\ph\> = \exp[\Om^{1/2}(\ph \ah^{\dg}_\bO - \ph^*
\ah_\bO)]|0\>.
 \ee
Once again, we have a set of degenerate ground states
labelled by a phase angle $\vt$.  None of these states
has a definite particle number.  However, as
$\Om\to\infty$ the uncertainty in $n_\bO$ tends to
zero.  Moreover, as before, the scalar product between
states with different values of $\vt$ tends to zero; one
finds
 \be
\<\et e^{i\vt}|\et e^{i\ch}\> =
\exp[-\Om\et^2(1-e^{-i(\vt-\ch)})] 
\to 0 \quad{\rm as}\quad\Om\to \infty.
 \ee
Here too, these states belong to distinct orthogonal
Hilbert spaces, carrying unitarily inequivalent
representations of the canonical commutation relations. 

It is important to note that degenerate ground states
and symmetry breaking occur only in the infinite-volume
limit.  In a finite volume, there is always a unique
ground state, a uniform superposition of `ground states'
with different phases.  As in the case of the
ferromagnet, the limits of infinite volume and zero
external field do not commute.  If we keep $j$ finite
and let $\Om\to\infty$, we arrive at a state with
$\<\phh(\br)\> = \et e^{i\vt}$.  On the other hand if we
let $j\to 0$ first, the particle number remains
definite, and we end up with a superposition of all the
degenerate ground states.  Physically (unless
$j\ne0$) the states are indistinguishable.

A Bose gas with interaction is described by the
Hamiltonian
 \ba
\Hh \!\!&=&\!\! \int_\Om d^3\br\;
{\hb^2\ov2m}\bna\phh\dg(\br)\cd\bna\phh(\br) \nonumber\\
&&+ {1\ov2} \int_\Om d^3\br \int_\Om d^3\br'\;
V(|\br'-\br|)\phh\dg(\br')\phh\dg(\br)\phh(\br)\phh(\br').  
 \ea
which is clearly still invariant under the
phase transformations (\ref{phaserot}).  So long as the
interaction is weak, the qualitative picture is largely
unchanged.  Bose-Einstein condensation has been
observed in alkali metal vapours, such as rubidium
\cite{And+95}.

\subsection{Liquid helium-4}

The normal-to-superfluid `lambda transition' at a
temperature of about 2 K in liquid $^4$He is another
example of a symmetry-breaking phase transitions in a
bosonic system \cite{TilT90}.  But unlike the case of
alkali metal vapours, this is a system with strong
interparticle interactions that substantially change the
picture.  Nevertheless, there are close similarities. 
The broken symmetry is still the phase symmetry
associated with conservation of particle number; the
transition may still be described in terms of a scalar
field
$\phh$ which acquires a non-zero expectation value below
the transition.

The most obvious difference caused by the interatomic
interactions is in the nature of the excitations above
the ground state.  In $^4$He at low temperature, these
are collective excitations, {\em phonons}, whose
dispersion relation is linear near $\bk=\bO$:
 \be
\ep_\bk = \cs\hb|\bk|,
 \ee
where $c\rms{s}$ is the sound speed.  At larger values
of $|\bk|$ the graph curves downwards to a minimum;
excitations near the minimum are called {\em rotons}.

A useful description of $^4$He is provided by the
Ginzburg--Landau model \cite{TilT90}, which may also be
applied to other systems.  The starting point is to
consider the {\em free energy}, $F$ say, as a function
of the order parameter $\ph$.  At least in principle,
$F[\ph]$ may be calculated in the usual way from the
partition function $Z$, by restricting the sum over
states to those with a given expectation value of the
order parameter field, $\<\phh(\br)\>=\ph(\br)$.  In the
neighbourhood of the transition temperature, $F$ can be
expanded in powers of $\ph$:
 \ba
F[\ph] \!\!&=&\!\! \int_\Om d^3\br\;
{\hb^2\ov2m}\bna\ph^*(\br)\cd\bna\ph(\br) \nonumber\\
&&+ \int_\Om d^3\br \left[\al(T)|\ph(\br)|^2 + \half
\beta(T) |\ph(\br)|^4 + \cdots\right],
 \label{FGL}
 \ea
where the higher terms are usually unimportant, at least
for a qualitative description.  The coefficient
$\beta(T)$ is always positive and may usually be taken
to be constant.  At high temperature, $\al(T)$ is also
positive so that the minimum of the free energy occurs at
$\ph=0$.  At low temperature, however, $\al$ becomes
negative, and the minimum occurs at
 \be
|\ph| = \left(-\al(T)\ov\beta(T)\right)^{1/2}.
 \ee
As usual, the phase of $\ph$ is arbitrary: we have a
degenerate equilibrium state.

Note that the critical temperature $\Tc$ is the
temperature at which $\al(T)$ $=0$.  Close to that point,
we may take 
 \be
\al(T) \ap \al_1\cd(T-\Tc).
 \ee

A good qualitative picture of the behaviour of $^4$He is
given by the two-fluid model, normal plus superfluid. 
The scalar field $\ph=\<\phh\>$ describes the superfluid
component, defining both the superfluid density and
velocity: 
 \be
\ns = |\ph|^2, \quad \bvs = {\hb\ov m}\bna\vt, \quad {\rm
where} \quad \ph=|\ph|e^{i\vt}.
 \label{He4vs}
 \ee
The normal component corresponds to single-particle (or,
rather, single-quasiparticle) excitations above the
ground state.

\subsection{Superconductors}

The electrons in a solid constitute a Fermi gas rather
than a Bose gas.  It is not single electrons that
condense but bound pairs of electrons, {\em Cooper
pairs} \cite{TilT90}.  There is an effective attractive
force between electrons near the Fermi surface
$\bk^2=\kF^2$.  At least for conventional
superconductors, this force preferentially binds pairs
with equal and opposite momenta and spins.

Below the critical temperature, we find that in the
ground state  
 \be
\<\ah_{\bk\ua}\ah_{-\bk\da}\> = F(k) \ne 0, \quad {\rm
for}\quad k=|\bk|\sim\kF.
 \label{scpairing}
 \ee
The order parameter $\phh$ in this case can be taken to
be an integral over such products of pairs of
destruction operators, multiplied by the internal wave
function of a Cooper pair.

There is an important difference between this and the
examples discussed previously.  The symmetry here is
again the phase symmetry (\ref{phaserot}), but it is now
a local, {\em gauge} symmetry: $\vt$ is allowed to be a
function of space and time, $\vt(t,\br)$.  This is
possible because of the coupling to the electromagnetic
field $A_\mu(x)$ which transforms as 
 \be
A_\mu(x) = A_\mu(x)-{{\hb\ov2e}}\pa_\mu\vt(x).
 \ee
The factor of $2e$ in the denominator appears because
this is the charge of a Cooper pair.  It ensures that
the {\em covariant derivative}
 \be
D_\mu\phh \equiv
\pa_\mu\phh+2i{e\ov\hb}A_\mu\phh,
 \ee
transforms in the same way as $\phh$ itself.  The
Ginzburg--Landau model may be used for superconductors
too, provided that the derivatives $\bna\ph$ in
(\ref{FGL}) are replaced by covariant derivatives
$\bD\ph$.

Symmetry breaking in gauge theories is a somewhat
problematic concept.  Indeed Elitzur's theorem
\cite{Eli75,ItzD89} says that {\em spontaneous breaking
of a local, gauge symmetry is impossible!} --- which
might be thought to imply that what I have just told you
is nonsense.  More specifically, it says that, while for
a global symmetry taking the infinite-volume limit and
then letting $j\to0$ may yield a state with
$\<\phh\>\ne0$, in a gauge theory we always have
 \be
\lim_{j\to0} \lim_{\Om\to\infty} \<\phh\> = 0.
 \ee 
But one must be careful not to misinterpret this
(entirely correct) theorem.  It applies {\em only} in an
explictly gauge-invariant formalism.  If, as is often
done, we add a gauge-fixing term that explictly breaks
the local symmetry (\eg, by imposing the Coulomb gauge
condition $\na\cd\bA=0$) then the remaining global
symmetry {\em can} be broken spontaneously.  We
certainly can define and use gauge-non-invariant states
with $\<\phh\>\ne0$, though there must always be an
alternative (but often inconvenient) gauge-invariant
description.

A model widely used as an exemplar of symmetry breaking
in particle physics is the Abelian Higgs model, the
relativistic version of the Ginzburg--Landau model.  It
is described by the action integral
 \be
I=\int d^4x\left[-\case{1}{4}F_{\mu\nu}F^{\mu\nu}
+D_\mu\ph^*D^\mu\ph
-\half\la(\ph^*\ph-\et^2)^2\right],
 \label{AHac}
 \ee
with
 \be
D_\mu\ph=\pa_\mu\ph+ieA_\mu\ph, \qquad 
F_{\mu\nu}=\pa_\mu A_\nu-\pa_\nu A_\mu.
 \ee
Here the coupling constant $\la$ plays the role of
$\beta$ and $-\la\et^2$ that of $\al$.  (Note that
here, and in general when dealing with relativistic
models, I set $c=\hb=1$.)  Symmetry breaking in this
model is very similar to that in a superconductor.

\subsection{Liquid crystals}

A very different example is provided by the isotropic to
nematic transition in a liquid crystal \cite{deGP93}.  A
nematic liquid crystal is typically composed of
rod-shaped molecules that  like to line up parallel to
one another.  There is no long-range translational order:
the molecules are free to flow past one another.  But
there is long-range orientational order.  At any point
in the liquid there is a preferred direction,
characterized by a unit vector $\bn$, the {\em
director}.  Note that $\bn$ and $-\bn$ are completely
equivalent.

The symmetry group here is the rotation group $SO(3)$. 
Above the transition temperature $\Tc$, the system is
completely isotropic, with the molecules randomly
oriented, but below it, the rotational symmetry is
broken. 

A convenient choice for the order parameter in this case
is the average mass quadrupole tensor $\bQ$ of the
molecules in a small region.  When the directions of the
molecules are isotropically distributed, $\bQ=\bO$.  But
if they are aligned in the direction of $\bn$, it has
the form
 \be
\bQ = Q(3\bn\bn-\bone).
 \ee
In particular, if $\bn$ is in the $z$ direction, then
 \be
\bQ = \pmat{-Q&0&0\cr 0&-Q&0\cr 0&0&2Q}.
 \ee

\subsection{Generic case}

Let us now examine the generic situation.  (For more
detail, see for example \cite{Kib00}.)  Suppose the
system has a symmetry group $G$.  In other words, the
Hamiltonian $\Hh$ is invariant under every operation
$g\in G$:
 \be 
\Uh^{-1}(g)\Hh\Uh(g) = \Hh \quad {\rm for\; all} \quad
g\in G, 
 \label{invH}
 \ee
where $\Uh(g)$ is the unitary operator representing the
operation $g$ on the Hilbert space.

However, we assume also that there is an operator $\phh$
with a nonvanishing ground-state expectation value which
transforms non-trivially under $G$.  Specifically, we
consider a multiplet of operators
$\phh=(\phh_i)_{i=1\dots n}$ transforming according to
some $n$-dimensional representation $D$ of
$G$:
 \be
\Uh^{-1}(g)\phh_i\Uh(g) = \sum_j D_{ij}(g)\phh_j,
 \ee
or more concisely
 \be
\Uh^{-1}(g)\phh\Uh(g) = D(g)\phh.
 \ee

We suppose that the expectation value in the ground
state $|0\>$,
 \be
\<0|\phh|0\> = \ph_0,
 \ee
say, is not invariant:
 \be 
\<0|\Uh^{-1}(g)\phh\Uh(g)|0\>=D(g)\ph_0 \ne \ph_0,
 \ee
for some $g\in G$.  Obviously this implies that the
ground state $|0\>$ is not invariant:
 \be 
\Uh(g)|0\> \ne |0\>.
 \ee
But by (\ref{invH}), $\Uh(g)|0\>$ is also an eigenstate
of $\Hh$ with the same eigenvalue; the ground state is
degenerate.

In general, not all elements of $G$ lead to distinct
ground states.  There may be some subgroup $H$ of
elements such that
 \be
D(h)\ph_0 = \ph_0 \quad {\rm for\; all} \quad h\in H.
 \ee
The distinct degenerate ground states correspond to the
distinct values of $\ph=D(g)\ph_0$.  Hence they are in
one-to-one correspondence with the {\em left cosets} of
$H$ in $G$ (sets of elements of the form $gH$).  These
cosets are the elements of the {\em quotient space}
 \be
M = G/H.
 \ee
This space may be regarded as the vacuum manifold or
manifold of degenerate ground states.

For example, for a Heisenberg ferromagnet, $G=SU(2)$,
and $H=U(1)$, the subgroup of rotations about the
direction of the magnetization vector.  Here $M=
SU(2)/U(1) = S^2$, a two-sphere.  For a Bose gas, $G=
U(1)$, and $H$ comprises the identity element only,
$H=\bone\equiv\{1\}$.  Thus $M = S^1$, the circle.

A nematic is a slightly less trivial example.  Here
$G=SO(3)$; however, $H$ is not merely the subgroup
$SO(2)\equiv C_\infty$ of rotations about $\bn$. 
Rather, $H$ is the infinite dihedral group, $H=D_\infty$,
which includes also rotations through $\pi$ about axes
perpendicular to $\bn$.  Correspondingly $M$ is not the
two-sphere but the real projective space $RP^2$,
obtained from $S^2$ by identification of opposite points.

\subsection{Helium-3}

Finally let me turn to the particularly interesting, and
relatively complicated, case of $^3$He
\cite{VolW90,Vol92}.  This lighter isotope also exhibits
a phase transition, though at a much lower temperature
than $^4$He, between 2 and 3 mK.  It is of course a {\em
Fermi} liquid.  So the mechanism of superfluidity is
very different, similar to that of superconductivity. 
In this case we have Cooper pairs not of electrons but
of $^3$He atoms.  The order parameter can again be
constructed from {\em pairs} of destruction operators.

There is however an important difference.  In the
original BCS model, the pairs were bound in an isotropic
$^1$S state; as indicated by the form of
(\ref{scpairing}).  But for a pair of
$^3$He atoms close to the Fermi surface it turns out
that the most attractive state is the $^3$P.  The pairs
have both unit orbital and unit spin angular momenta:
$L=S=1$.  We need to consider a more general form of
order parameter, related to the quantity
 \be
F_{ab}(\bk) = \<\ah_{\bk a}\ah_{-\bk b}\>, \qquad a,b =
\ua,\da.
 \ee
The fact that $S=1$ tells us that $F$ should be
symmetric in the spin indices $a$ and $b$, so it can be
expanded in terms of the three independent symmetric
$2\X2$ matrices, $\si_j i\si_2$, where $\si_j$ are the
Pauli matrices.  The fact that $L=1$ means that $F$
should be proportional to $\bk$ times a function of
$k=|\bk|$ only.  Thus we can write
 \be
F_{ab}(\bk) = F(k)A_{ij}(\si_i i\si_2)_{ab}k_j,
 \ee
where the two-index tensor $A$ may be normalized by
$\tr(A\dg A)=1$.

The order parameter is essentially $A$ times a scalar
factor representing the density of Cooper pairs.  Since
it is now a $3\X3$ complex matrix rather than a scalar,
the possible patterns of symmetry breaking are much more
complex.  There are in fact two distinct superfluid
phases, $^3$He-$A$ and $^3$He-$B$, which are stable in
different regions of the phase diagram; the $A$ phase
is stable only at high pressure and at temperatures not
far below the critical temperature.  In the
presence of a magnetic field, there is a third stable
phase, the $A1$ phase.

The system exhibits a much larger symmetry than $^4$He. 
To a good approximation, it is symmetric under {\em
independent} orbital and spin rotations, as well as
under the phase rotations as before.  Thus the symmetry
group is 
 \be
G=SO(3)_\bS\X SO(3)_\bL\X U(1)
 \ee
(There is also a weak spin-orbit coupling, whose effects
I will discuss a little later.)

In the $A$ phase, the order parameter takes the form
 \be
A_{ij} = {1\ov\sqrt2}d_i (m_j + in_j),
 \ee
where $\bd, \bm$ and $\bn$ are unit vectors, with
$\bm\perp\bn$.  The vector $\bd$ defines an axis along
which the component of $\bS$ vanishes.  If we define
$\bl=\bm\wedge\bn$, then $\bl$ is an axis along which
the component of $\bL$ is $+1$.  In this case, the
subgroup $H$ that leaves $A$ invariant comprises
spin rotations about the direction of $\bd$, orbital
rotations about $\bl$ {\em combined} with compensating
phase transformations, and, finally, the discrete
transformation that reverses the signs of all three
vectors.  Hence 
 \be 
H_A=U(1)\X U(1)\X \bZ_2.
 \ee
Correspondingly, the vacuum manifold is 
 \be
M_A = G/H_A = S^2\X SO(3)/\bZ_2.
 \label{MHe3A}
 \ee
Here, the elements of $S^2$ label the direction of
$\bd$, while $SO(3)$ describes the orientation of the
orthonormal triad $(\bl,\bm,\bn)$.  The $\bZ_2$ factor
represents the identification
$(\bd,\bm,\bn)\equiv(-\bd,-\bm,-\bn)$.

The $B$ phase, by contrast, is characterized by an order
parameter of the form
 \be
A_{ij} = R_{ij}e^{i\vt},
\label{Bordpar}
 \ee
where $R\in SO(3)$ is a real, orthogonal matrix.  In
this case, the only elements of $H$ are {\em combined}
orbital and spin rotations, so
 \be
H_B = SO(3) \qquad {\rm and} \qquad
M_B = G/H_B = SO(3)\X S^1.
 \label{MHe3B}
 \ee
   
As I mentioned earlier, there is actually a weak
spin-orbit coupling term in the Hamiltonian, which is
only noticeable at long range, and which reduces the
symmetry to
 \be
G'=SO(3)_\bJ\X U(1) \quad {\rm with} \quad \bJ=\bL+\bS. 
 \ee
Note that going from $G$ to $G'$ is not strictly speaking
a case of spontaneous symmetry breaking.  There are
similarities: at short range, the symmetry appears to be
the larger group $G$; when we go to long range (or low
energy), we see that the symmetry group is actually
$G'$.  However, the true symmetry is always $G'$; $G$ is
only approximate. 

In the $A$ phase the effect is to require that the
vectors $\bd$ and $\bl$ be parallel or antiparallel, and
in fact by selecting one of the two configurations
related by inversion, we can ensure that $\bd=\bl$.  In
this case, we find
 \be
H'_A = U(1), \qquad M'_A = G'/H'_A = SO(3).
 \label{MpHe3A}
 \ee
In the $B$ phase, the restriction is that $R$ in
(\ref{Bordpar}) is no longer an unconstrained orthogonal
matrix, but a rotation matrix through a definite angle
(the {\em Leggett angle} $\th\rms{L}=\arccos(-{1\ov4})$)
about an arbitrary axis $\bn$.  Thus we find
 \be
H'_B = SO(2), \qquad M'_B = G'/H'_B = S^2 \X S^1.
 \label{MpHe3B}
 \ee

\subsection{The standard model of particle physics}

There are remarkable similarities between the symmetry
breaking pattern of $^3$He and that found in the standard
model of particle physics which incorporates quantum
chromodynamics together with the unified electroweak
theory of Weinberg and Salam.  It is based on the symmetry
group
 \be
G=SU(3)\rms{col}\X SU(2)_\bI\X U(1)_Y,
 \label{Gsm}
 \ee 
where $\bI$ and $Y$ denote respectively the weak isospin
and weak hypercharge.  The symmetry breaking from this
down to the observed low energy symmetry is described by
the Higgs field, $\phh$, which plays the role of the order
parameter.  It is a two-component complex scalar field
invariant under the colour group $SU(3)\rms{col}$,
belonging to the fundamental 2-dimensional representation
of $SU(2)_\bI$, and with non-zero weak hypercharge
$Y=1$.  It acquires a vacuum expectation value of the
form
 \be
\<\phh\> = \ph = \pmat{0\cr v},
 \ee
thus reducing the symmetry to the subgroup
 \be
H=SU(3)\rms{col}\X U(1)\rms{em}.
 \ee
The generator of the remaining $U(1)$ symmetry is the
electromagnetic charge 
 \be
Q=I_3+\half Y.
 \ee

There may also be other stages of symmetry breaking at
higher energies.  The three independent coupling constants
$g_3, g_2, g_1$ corresponding to the three factors in $G$
have a weak logarithmic energy dependence and appear to
come to approximately the same value at an energy scale
of about $10^{15}$ GeV, especially if supersymmetry is
incorporated into the model \cite{Ama+91,Ama+92,Hab98}. 
This suggests that there may be a {\em grand unified
theory} (GUT) uniting the strong, weak and
electromagnetic interactions in a single theory with a
symmetry group such as $SO(10)$.  There would then be a
phase transition (or a sequence of phase transitions) at
that energy scale at which the GUT symmetry breaks to the
symmetry group (\ref{Gsm}) of the standard model.  If
the model is supersymmetric, then there must also be a
supersymmetry-breaking transition.

\section{Defect formation}

The appearance of topological defects is a common feature
of symmetry-breaking phase transitions.  In this lecture,
I shall review the defects associated with the various
transitions discussed earlier, and the general conditions
for the existence of defects.

\subsection{Discrete symmetry breaking}

The simplest possible field-theoretic model that exhibits
symmetry breaking is a model of a {\em real} scalar field
described by the action integral
 \be
I = \int d^4x\left[\half(\pa_\mu\ph)(\pa^\mu\ph) -
\case{1}{8}\la(\ph^2-\et^2)^2\right].
 \ee
Here the action is invariant under the reflection
symmetry $\ph\mapsto-\ph$.  Thus the symmetry group is
$G=\bZ_2$, and the manifold of degenerate vacua reduces
to a pair of points; the two vacuum states are
characterized by
 \be
\ph \equiv \<\phh\> = \pm \et.
 \ee

At high temperature, the equilibrium state is symmetric,
with $\<\phh\>=0$.  When the system cools through the
critical temperature, $\phh$ acquires a non-zero
expectation value, but the sign is chosen arbitrarily. 
So it may happen that in one region, it chooses
$\ph\ap\et$ and in another $\ph\ap-\et$.  When
such regions meet, they must be separated by a planar
defect, a {\em domain wall}, across which $\ph$ goes
smoothly from one value to the other.  The minimum energy
configuration is determined by a balance between gradient
energy and potential energy.  At zero temperature one
finds for example that a domain wall in the
$xy$-plane is described by
 \be
\ph(z) = \et\tanh{z\ov \xi}, \quad {\rm with} \quad
\xi={2\ov\sqrt\la\et}.
 \ee

As the system cools below $\Tc$, energy is trapped in
the domain wall.  In a sense the defect is a region
of trapped old high-temperature phase, with the
characteristic energy density that it had at $\Tc$.  The
wall is {\em topologically stable}.  It can move, as one
domain grows at the expense of the other, but it cannot
simply break.  A closed wall bounding a finite domain
may of course shrink and eventually disappear.  But this
is a relatively slow process, so walls may have
continuing effects.

\subsection{Abelian vortices or strings}

Now let us consider the case of an Abelian $U(1)$
symmetry, such as that of superfluid $^4$He.  When the
system is cooled through the transition temperature, the
order parameter acquires a non-zero expectation value
$\ph=\et e^{i\vt}$.  The magnitude $\et$ is determined by
the minimization of the free energy, but the phase $\vt$
is arbitrary.  It is chosen randomly.  However, in a large
system there is no reason why the same choice should be
made everywhere; $\vt$ may vary from one part of the
system to another.  We should expect the choice to be made
independently in widely separated regions, especially if
we are talking about a transition in the early Universe,
where such regions may have had no prior causal contact.

When such a random choice is made, it may happen that
around some large loop in space the value of $\vt$ varies
through $2\pi$ or a multiple thereof.  In such a case,
$\ph$ must vanish somewhere inside the loop; indeed it
must vanish all along a curve that threads through the
loop.  This is the core of a topological defect, a cosmic
string or vortex.

If the string is along the $z$ axis, the order parameter
around it typically takes the form
 \be
\ph(r,\vp,z) = \et f(r)e^{in\vp},
 \label{stringphi}
 \ee
where $r,\vp,z$ are cylindrical polar coordinates, and $n$
is an integer, the {\em winding number}.  The function
$f$ has limiting values $f(0)=0$, $f(\infty)=1$.  It may
be determined by minimizing the Ginzburg--Landau free
energy, (\ref{FGL}).

For a superfluid, an important consequence of the
expression (\ref{He4vs}) for the superfluid velocity is
that the superfluid flow is irrotational:
$\bna\wedge\bvs=\bO$.  The vorticity vanishes everywhere,
{\em except} in the core of the string, where the
superfluid density vanishes.  The string is a vortex.  The
form of (\ref{stringphi}) implies that there is a flow of
superfluid around the string, with velocity
 \be
(\vs)_\vp = {\hb\ov m_4r}
 \ee
at large $r$, where $m_4$ is the mass of a $^4$He atom. 
Thus the {\em circulation} around the string is quantized:
 \be
\oint \bvs\cd d\br = n\ka_4,
 \ee
where the circulation quantum is
 \be
\ka_4 = {2\pi\hb\ov m_4}.
 \ee

There is a similar vortex in $^3$He-$B$, but in that case
$m_4$ is replaced by the mass of a Cooper pair, namely
$2m_3$, so the circulation quantum is 
 \be
\ka_3 = {\pi\hb\ov m_3}.
 \ee

An important feature of the string or vortex is its
topological stability deriving from this quantization. 
It can move around, but cannot break.  A vortex
loop can disappear by shrinking to a point, but a long,
straight vortex is stable.

\subsection{Vortices in a gauge theory}

If the symmetry is a gauge symmetry, with coupling to a
gauge field $A_\mu$, then around the string $A$ has an
azimuthal component, 
 \be
A_\vp(r,\vp,z) = {n\hb\ov er}g(r),
 \ee
where $e$ is the charge and $g$ has the same limiting
values as $f$.  By taking an integral round a large loop
surrounding the string we find that it carries a
quantized magnetic flux,
 \be
\Ph = \lim_{r\to\infty} \int_0^{2\pi} A_\vp r d\vp =
n{2\pi\hb\ov e}.  
 \ee
The magnetic field is given by
 \be
B_z ={n\hb\ov er}g'(r).
 \ee
(In the case of a superconductor $e$ in the above should
be replaced by the charge $2e$ of a Cooper pair, so the
flux quantum is actually $\pi\hb/e$.)

The functions $f$ and $g$ are determined by minimizing
the free energy.  For the Abelian Higgs model at zero
temperature, with action integral (\ref{AHac}) (setting
$c=\hb=1$), they satisfy the equations
 \ba
f''+{1\ov r}f'-{n^2\ov r^2}(1-g)^2f+\la\et^2(1-f^2)f
\!\!&=&\!\! 0,
\nonumber\\ 
g''-{1\ov r}g'+2e^2\et^2f^2(1-g) \!\!&=&\!\! 0.
 \ea
No analytic solution is known, but it is easy to find
solutions numerically.  Note that there are two length
scales governing the large-$r$ behaviour of the
functions, the inverse masses of the scalar and vector
excitations (Higgs and gauge particles),
 \be
\ms^2=2\la\et^2, \quad {\rm and} \quad
\mv^2=2e^2\et^2.
 \ee
The asymptotic behaviour depends on the ratio of
these two,
 \be
\ka={\ms^2\ov\mv^2}={\la\ov e^2}.
 \ee
For $\ka<4$ one finds that at large $r$ 
 \be
1-g\propto r^{1/2}e^{-\mv r}, \qquad 
1-f\propto r^{-1/2}e^{-\ms r}.
 \ee
In this case, the string has a narrow core that
constitutes a magnetic flux tube, while the order
parameter reaches its vacuum value over a larger
distance.  On the other hand, when $\ka>4$, $\mv$
controls the behaviour of both $1-f$ and $1-g$, with
 \be
1-g\propto r^{1/2}e^{-\mv r}, \qquad 
1-f\propto r^{-1}e^{-2\mv r}.
 \ee

In superconductors \cite{TilT90}, the two length scales
are known as  the correlation length $\xi=\hb/\ms c$ and
the Landau penetration depth $\la=\hb/\mv c$.  Here large
and small values of $\ka$ distinguish so-called type-I
from type-II superconductors.  In a type-II
superconductor, vortices with $|n|>1$ are unstable;
there is a repulsive force between parallel $n=1$
vortices which can stabilize a lattice of vortices. 
Hence there is an intermediate range of magnetic field
strength within which the field penetrates the
superconductor but confined to a lattice of flux tubes.

\subsection{Defects in nematics}

It is easy to construct a stable linear defect similar to
a string in the case of a nematic, by allowing the
director $\bn$ to rotate as one moves around the string
through an angle $\pi$: \eg, we can take
 \be
\bn(r,\vp,z) = (\cos{\vp\ov2},\sin{\vp\ov2},0),
 \ee
provided we include in the order parameter $\bQ$ a factor
$f(r)$ that vanishes at $r=0$.  $\bQ$ then has no
singularity because of the identification of $\bn$ and
$-\bn$.  This configuration is called a {\em
disclination} \cite{deGP93}.

Like the superfluid vortex, the disclination is
topologically stable.  If the rotation angle were $2\pi$
instead of $\pi$ it would not be.  It could `escape into
the third dimension': at small $r$, we could rotate $\bn$
upwards until at $r=0$ it points in the $z$ direction,
thus allowing $f(0)$ to be nonvanishing.

In addition to this linear defect there can be a point
defect in a nematic liquid.  Away from the centre,
$\br=\bO$ say, we can take
 \be
\bn(\br) = {\br\ov r},
 \ee
again provided there is a factor in $\bQ$ that vanishes
at the centre.

This is often called the {\em hedgehog} or {\em monopole}
configuration.  Like the vortex it is topologically
stable, and cannot disappear spontaneously --- though it
may annihilate with an anti-hedgehog.

\subsection{The fundamental group}

The general conditions for the existence of defects can
be expressed in terms of the topology of the vacuum
manifold $M$, specifically its {\em homotopy groups}
\cite{Hu59}.  

The existence of linear defects for example is related
to the possibility of finding non-trivial closed loops in
$M$.  Let us recall that different points in $M$
correspond to different values of the order parameter
labelling different degenerate vacua.  A closed loop is a
continuous map $\ph: I\to M$ from the unit interval of
real numbers, $I=[0,1]\subset \bR$, to $M$ such that
$\ph(0)=\ph(1)$ (or equivalently a map from the circle
$S^1$ to $M$).  Linear defects can exist if it is
possible to find a closed loop that cannot be
continuously shrunk to a point without leaving $M$,
because then if the value of the order parameter around a
loop in space follow this curve, it is not possible to
fill in the values inside the loop continuously while
remaining on $M$.

In general, two closed loops are {\em homotopic} if it is
possible to deform one continuously into the other
within $M$.  This is an equivalence relation, so we may
define {\em homotopy classes} of loops.  For example,
when $M$ is the circle $S^1$, the homotopy classes may be
labelled by the {\em winding number}, the (algebraic)
number of times that we traverse the circle while going
from $0$ to $1$ in $I$.

The homotopy classes constitute the elements of a group,
the {\em fundamental group} or {\em first homotopy group}
of $M$, denoted by $\pi_1(M)$.  To construct it, we
introduce a base point $b\in M$, and consider loops
starting and finishing at $b$, \ie\ $\ph(0)=\ph(1)=b$. 
Then the product of two loops $\ph$ and $\ps$ is the loop
constructed by following $\ph$ and then $\ps$:
 \be
(\ph\cd\ps)(t)=
\cases{\ph(2t)\quad {\rm for}\quad t\le\half,\cr
\ps(2t-1)\quad {\rm for}\quad t>\half.}
 \ee
It is easy to show that this is a relation between
homotopy classes, and so defines a product on the set of
classes, which thus becomes the group $\pi_1(M)$.  The
condition for the existence of topologically stable
linear defects, strings or vortices, is that the
fundamental group be non-trivial: $\pi_1(M)\ne\bone$.

For the Abelian case, where $M=S^1$, the fundamental group
is simply the group of integers, $\pi_1(S^1)=\bZ$.  The
distinct possible linear defects are labelled by the
elements of this group, the winding numbers.

In a continuum version of the Heisenberg ferromagnet, we
have $M=S^2$.  On the sphere, all loops can be shrunk to
a point, so $\pi_1(S^2)=\bone$; there are no possible
linear defects.

For the nematic, however, the situation is different.  On
the sphere $S^2$ all loops are homotopically trivial, but
this is no longer true when we identify opposite points to
form $RP^2$, because a curve that starts at one pole and
ends at the opposite pole is closed in $RP^2$ but cannot
be shrunk to a point.  There is only one non-trivial
homotopy class, because traversing the same loop twice
gives a trivial loop; as we noted earlier, a disclination
with a winding of $2\pi$ rather than $\pi$ is not
stable.  Hence the fundamental group in this case is
$\pi_1(RP^2)=\bZ_2=\{0,1\}$, the group of integers modulo
2.

\subsection{The second homotopy group}

The conditions for the existence of other types of
defects can also be expressed in terms of homotopy
groups.  For point defects such as the hedgehog the
relevant question is whether there are non-shrinkable
two-surfaces in $M$.

The second homotopy group \cite{Hu59} is defined in
terms of closed two-surfaces, \ie\ maps $\ph:I^2\to M$
from the unit square to $M$, such that
 \be
\ph(0,t)=\ph(1,t)=b, \quad
\ph(s,0)=\ph(s,1)=b \quad 
{\rm for\;all}\; s \;{\rm and}\; t.  
 \ee
In other words, $\ph$ maps the whole boundary of $I^2$ to
the chosen base point $b\in M$.  In effect, it is a map
from $S^2$ to $M$, in which one designated point is
mapped to $b$.

Two closed surfaces are {\em homotopic} if one can be
smoothly deformed into the other.  This defines an
equivalence relation, and hence a classification into
homotopy classes of surfaces.

As before, we can introduce a product on the set of
closed surfaces, by setting
 \be
(\ph\cd\ps)(s,t)=
\cases{\ph(s,2t)\quad {\rm for}\quad t\le\half,\cr
\ps(s,2t-1)\quad {\rm for}\quad t>\half.}
 \ee
This defines a product on the set of homotopy classes. 
(We could equally well have defined the product with the
roles of $s$ and $t$ interchanged; it is easy to show
that the results are homotopic to each other.)  Thus we
have defined the {\em second homotopy group} $\pi_2(M)$.
 
As a simple example, let us consider a continuum version
of the Heisenberg ferromagnet, with $SU(2)$ symmetry and
an order parameter $\bM$ transforming according
to the 3-dimensional vector representation.  Below the
transition, the magnitude of $\bM$ is fixed but its
direction is arbitrary.  Thus the subgroup
$H$ that leaves $\bM$ invariant is $H=U(1)$ and
$M=SU(2)/U(1)$ $=S^2$.  In this case, the homotopy
classes are labelled by an integer, the (algebraic)
number of times the map wraps around the sphere.  For
example, a typical element of the homotopy class
labelled by $n$ is the map of one sphere on another
defined in terms of polar coordinates $\th,\vp$ by
 \be
\ph: S^2\to S^2: (\th,\vp)\mapsto(\th,n\vp).
 \ee
For this case, therefore, $\pi_2(S^2)=\bZ$, the group of
integers.

For the nematic, we have to identify opposite points of
$S^2$ and pass to $RP^2$ but this makes no difference to
the classification of closed surfaces.  We again have
$\pi_2(RP^2)=\bZ$, so the possible hedgehogs are labelled
by an integer.

One thing this classification cannot tell us, however, is
whether configurations with $|n|>1$ are actually stable. 
In some cases, it may be energetically favourable for a
configuration with winding number $n=2$ for instance, to
break up into two separate $n=1$ configurations.  Whether
this actually happens is a question of detailed dynamics.

\subsection{Domain walls}

As we noted earlier, domain walls occur when a {\em
discrete} symmetry is broken.  More generally, the
condition for the existence of domain walls is that the
vacuum manifold $M$ be disconnected.  Domain walls are
classified by the elements of what is often called the
`{\em zeroth homotopy group}', denoted by $\pi_0(M)$,
whose elements are in one-to-one correspondence with the
connected components of $M$.  It is analogous to the
higher homotopy groups: it may be regarded as
classifying maps $\ph: S^0\to M$, where $S^0$ is the
0-sphere (the boundary of the interval
$[-1,1]\subset\bR$, namely the pair of points
$\{1,-1\}$) in which the image of one chosen point is
the base point of $M$.

In a general case, the terminology is strictly speaking
inaccurate, because $\pi_0(M)$ is not a group.  There is
one special case in which it is so, namely when $H=\bone$,
so that $M$ is itself a group, $M=G$.  In this case, the
connected component $G_0$ of $G$ containing the identity
is an invariant subgroup (\ie, for any $g\in G$,
$gG_0g^{-1}=G_0$), and hence the quotient group $G/G_0$
is defined; moreover
 \be
\pi_0(G) = G/G_0.
 \ee 

Another way of characterizing $\pi_0(M)$ is as a
quotient of two groups.  If $\ph\in M$ can be connected
by a continuous path to $\ph_0$, then one can always find
a continuous path in $G$ from the identity $e$ to $g$
such that $D(g)\ph_0=\ph$.  Hence the connected
component $M_0\subset M$ containing $\ph_0$ may be
identified with the set of elements $\{g\ph_0: g\in
G_0\}$.  The subgroup of $G_0$ which leaves $\ph_0$
unaltered is clearly $H\cap G_0$.  Hence,
 \be
M_0 = G_0/(H\cap G_0).
 \ee
Now since $G_0$ is an invariant subgroup in $G$, it
follows that $H\cap G_0$ is also an invariant subgroup of
$H$.  Thus $H/(H\cap G_0)$ is a group, and moreover a
subgroup of $G/G_0$.  One can then show that
 \be
\pi_0(M) = (G/G_0)/(H/(H\cap G_0)).
 \ee

\subsection{Helium-3}

Finally, let us return to the interesting case of
$^3$He, beginning with the superfluid $^3$He-$B$ phase,
and initially ignoring the spin-orbit interaction.

We recall from (\ref{MHe3B}) that $M_B=SO(3)\X S^1$. 
Here both the zeroth and second homotopy groups
$\pi_0(M_B)$ and $\pi_2(M_B)$ are trivial, so there are
no topologically stable domain walls or monopoles. 
However,
 \be
\pi_1(M_B) = \bZ_2\X\bZ,
 \ee
so there are two different kinds of vortices.  The factor
$\bZ$ classifies vortices around which the phase changes
by $2n\pi$, exactly as in the case of $^4$He.  However,
the $\bZ_2$ factor arises because there are non-trivial
loops in $SO(3)=RP^3$.  In a vortex corresponding to the
non-trivial element of $\bZ_2$ there is no actual
circulation around the string, but rather a relative
rotation of the orbital and spin angular momenta.  These
are called {\em spin vortices} as opposed to {\em mass
vortices}.

Note that vortices may carry both types of quantum
number simultaneously.  Such a combination is a {\em
spin--mass vortex}.

When we take account of the spin-orbit interaction the
manifold is reduced, according to (\ref{MpHe3B}), to
$M'_B=S^2\X S^1$.   In this case, we find 
 \be
\pi_2(M'_B) = \bZ. 
 \ee 
Viewed on a large scale
there {\it are} monopole configurations.  But since there
are no short-range monopoles, these have no actual
singularity.  What happens is that the order parameter
near the monopole is forced to leave the manifold $M'_B$,
but can remain everywhere on the larger manifold $M_B$;
the rotation angle in the order parameter, which is fixed
to be the Leggett angle $\th\rms{L}$ at large distances,
can tend smoothly to zero at the centre, but $\ph$
itself remains non-zero.

We also find
 \be
\pi_1(M'_B) = \bZ,
 \ee
corresponding to the fact that the mass vortices are
unaffected by the spin-orbit coupling, and survive to
large distances.  This is not the case, however, for the
spin vortices, since there is no longer a $\bZ_2$
factor.  What happens is that these become attached to a
long-range soliton or domain-wall feature.  The order
parameter at large distances around this vortex cannot
lie everywhere on $M'_B$, but to minimize the energy it
does so except near one direction.  Note that this is
true in spite of the fact that $\pi_0(M'_B) = \bone$,
which means there are no truly stable domain walls:
$M'_B$ is connected, but  the relevant point is that it
is not possible in $M_B$ to deform the relevant loop in
such a way that it lies entirely in $M'_B$.  (Such cases
may be classified by the {\em relative} homotopy groups
of $M_B$ and its subspace $M'_B$, in this case the group
$\pi_1(M_B,M'_B)$.)

Now let us turn to the $A$ phase, for which according to
(\ref{MHe3A}) the `vacuum manifold' is $M_A=S^2\X
SO(3)/\bZ_2$.  This space is again connected, so there
are no stable domain walls.  However, we find 
 \be
\pi_1(M_A)=\bZ_4, \qquad \pi_2(M_A)=\bZ.
 \ee
Thus there are monopoles, labelled by an integer winding
number, and vortices labelled by an integer $n$ modulo
4.  On the other hand at long range the manifold,
given by (\ref{MpHe3A}), is simply $M'_A=SO(3)$, whence
 \be
\pi_1(M'_A)=\bZ_2, \qquad \pi_2(M'_A)=\bone.
 \ee
Hence there are no stable monopoles and only one class
of stable vortices; the latter are to be identified with
the $n=2$ short-range vortices.

It is not hard to see what happens to the other
short-range defects.  For an $n=\pm1$ short-range
vortex, the corresponding loop in $M_A$ cannot be
deformed to lie entirely within $M'_A$.  In other words,
we cannot make $\bd$ parallel to $\bl$ everywhere.  The
vortex becomes attached to a sheet or domain wall across
which $\bd$ rotates by $\pi$ with a compensating
rotation about $\bl$.  

Similarly, around a short-range monopole we cannot
deform the order parameter so that $\bd$ remains
everywhere parallel to $\bl$.  The monopole becomes
attached to a string in the centre of which $\bd$ is in
the opposite direction.

\section{Cosmology in the Laboratory}

Our present understanding of fundamental particle
physics suggests that the Universe may have undergone a
series of phase transitions very early in its history. 
One of the clearest signatures of these transitions
would be the formation of stable topological defects
with potentially significant cosmological effects.  To
predict these we need to estimate how many defects would
have been formed and how they would have evolved during
the subsequent cosmic expansion.  Calculations of the
behaviour of the system in the highly non-equilibrium
context of a rapid phase transition are problematic,
however, and it is hard to know whether they are
reliable.  There is no direct way of testing them,
because we cannot do experiments on the early Universe.

But what we {\em can} do is to apply similar methods to
analogous low-temperature transitions in
condensed-matter systems, which often have a very
similar mathematical description.  Over the last few
years, several experiments have been done in a variety
of systems to test ideas drawn from cosmology.  This has
led to some extremely innovative and exciting
condensed-matter physics, although the rsults are still
somehwat confusing.

\subsection{Defect formation in the early Universe}

The electroweak transition, at about $100$ GeV, where the
$W$ and $Z$ particles acquire a mass through the Higgs
mechanism, occurred when the age of the Universe was
around $10^{-10}$ s.  It is now believed that this is
not in fact a genuine phase transition but rather a
rapid but smooth crossover \cite{Kaj+96}.  (This is
possible only because this is a {\em gauge} theory.) 
There was probably a later transition, the quark--hadron
transition at which the soup of quarks and gluons
separated into individual hadrons.

More interesting from a cosmological point of view,
however, are the hypothetical transitions at even
earlier times.  If the idea of grand unification is
correct, there would have been a phase transition of
some kind at an energy scale of around $10^{15}$ GeV,
corresponding to a time about $10^{-36}$ s after the Big
Bang.  In some models, we expect a sequence of phase
transitions, as the symmetry is broken in several
stages, for example 
 \be
SO(10) \to SU(5) \to SU(3)\X SU(2)\X U(1),
 \ee
or     
 \be
SO(10) \to SU(4)\X SU(2)\X SU(2) \to SU(3)\X SU(2)\X
U(1). 
 \ee
The most attractive GUTs are supersymmetric.  Since
supersymmetry is not manifest at low energies, it must
have been broken at some intermediate time,  possibly
yielding another phase transition, perhaps at about 1
TeV. 

Domain walls, strings and monopoles may all have been
formed at early-Universe phase transtions, as indeed may
more general composite objects of various kinds
\cite{Kib00}.  Monopoles and domain walls are
cosmologically problematic, for different reasons. 
Heavy domain walls, such as those that could have been
formed in the early Universe, certainly do not exist in
our Universe today, and monopoles could be present only
in very small numbers.  So if these defects were produced
at all, there must have been a mechanism to remove them
completely or almost completely at an early stage. 
Inflation has often been invoked to do this job.

Strings on the other hand could have survived in
sufficient numbers to be cosmologically significant
without violating any observational bounds.  For a long
time it was believed that they might serve to explain
the initial inhomogeneities in the density of the
Universe from which galaxies and clusters later
evolved.  The idea that strings alone could seed these
density perturbations seems no longer viable, in the
light of the data on the cosmic microwave background
anisotropy.  It is still perfectly possible to fit the
data with models incorporating both strings and inflation
\cite{ConHM98}, but a recent analysis concludes that
strings probably do not make a significant contribution
\cite{Dur+02}.  They may, however, have had other
important cosmological effects, for example in the
generation of magnetic fields
\cite{VacV91,AveS95,Dim98}, high-energy cosmic rays
\cite{BonP97,BhaS00} and baryogenesis
\cite{DavP97,DimD99}.

For these reasons I shall restrict the discussion to the
case of string formation.  To be specific, let us
consider the breaking of an Abelian $U(1)$ symmetry ---
though most of the discussion can easily be extended to
non-Abelian symmetries.

\subsection{Defect formation at a first-order transition}

The nature of the early-Universe transitions is largely
unknown, in particular the order of each transition.  In
some cases, as I mentioned, there may be no true
transition at all.  Defects may be formed in any event,
but the mechanism depends strongly on the order.  In
relation to condensed-matter analogues, most interest
attaches to second-order transitions, and that is the
case I will spend most time on.  But I shall begin
with what is in some ways the simpler situation of a
first-order transition.

In fact, the first `cosmology in the laboratory'
experiments were done with a first-order transition,
namely the transition from normal isotropic liquid to
nematic liquid crystal \cite{Chu+91,Bow+94}.

Let us suppose, therefore, that there is a first-order
transition, proceeding by bubble nucleation.  Once the
Universe reaches the relevant critical temperature,
bubbles of the new low-temperature phase are born at
random positions and start to grow until they eventually
meet and merge.  The nucleation rate $\ga$ per unit
space-time volume is given by an expression of the form
 \be
\ga(T) = A(T) e^{-S\rms{E}(T)}
 \ee
where $S\rms{E}$ is the Euclidean action for a
tunnelling solution, and the prefactor $A$ is typically
of order $T^4$.  The nucleation rate determines the
characteristic distance $\xi$ between nucleation sites,
such that the number of separate bubbles nucleating in
a large volume $V$ is $V/\xi^3$.  Typically the bubbles
expand at relativistic speeds, and then $\xi$ is of order
$\ga^{-1/4}$.

In each new bubble the order parameter $\ph$ becomes
non-zero, and must choose a random phase $\vt$.  There
is no reason why there should be any correlation between
the phases in different bubbles (except conceivably in
the case of very near neighbours).  So it is reasonable
to assume that each is an independent random variable,
uniformly distributed between $0$ and $2\pi$.

When two bubbles meet, an equilibration process will
occur, leading to a phase $\vt$ smoothly interpolating
between $\vt_1$ and $\vt_2$ across the boundary.  It is
reasonable to assume that it will do so by the shortest
possible path, so that the total variation will always
be less than $\pi$; this is called the {\em geodesic
rule}.  Numerical simulations have confirmed that it is
usually though not universally true --- the rule may lead
to a slight underestimate of the total number of defects
formed \cite{Sri92,PogV98}.

When these two bubbles encounter a third, it is possible
that a string defect may be trapped along the line where
they meet.  This will happen if the net phase change
from $\vt_1$ to $\vt_2$ to $\vt_3$ and back to $\vt_1$
is $\pm2\pi$ rather than zero.  If the geodesic rule
applies and the three phases are strictly independent,
the probablility of this happening can easily be
seen to be $1\ov4$.

Thus the total length of string formed in this process
in a large volume $V$ will be of order $V/\xi^2$.  The
length of string per unit volume will be
 \be
L = {k\ov\xi^2},
 \label{Lxi}
 \ee
where $k$ is a numerical constant of order $1$.  (For
example, if it is assumed that nucleation sites form a
body-centred cubic lattice, one finds $k=3/2^{7/6}=1.34$.
A random lattice would really be more appropriate; that
might well give a somewhat smaller value.)

The first tests of this idea in condensed-matter systems
were done in nematic liquid crystals, by studying the
formation of disclination lines in the isotropic to
nematic transition \cite{Chu+91,Bow+94}.  The symmetry
in that case is of course non-Abelian, but the principle
is the same.  We may assume that within each nucleating
bubble of the nematic phase, the director $\bn$ is an
independent random variable, uniformly distributed over
half the unit sphere (except near the walls where special
effects come into play).

The analogue of the geodesic rule is then the assumption
that across the interface between two bubbles, the
director always turns by an angle less than $\pi$.  In
that case, the probability that a disclination will be
trapped between three bubbles with independently
oriented directors is $1/\pi$, so (\ref{Lxi}) should
still hold.

The experiments did in fact show reasonably good
agreement with the predictions.  Further experiments
have been done to check the correlations between defects
and antidefects \cite{Dig+99}. 

\subsection{Second-order transitions}

The argument is a little more complex in the case of a
second-order phase transition.  As the system cools
through the critical temperature, the order parameter
must acquire a non-zero value and choose a random
phase.  We may assume that the choice is made
independently in widely separated regions.  Thus there is
a chance that defects will be trapped, and we should
expect the formation of a random tangle of strings. 
What is less obvious is what the characteristic scale
$\xs$ of this tangle should be.  Here $\xs$ may be
defined by the condition that in a randomly chosen
volume $\xs^3$ there will be on average a length $\xs$
of string.  In other words, the length of string per unit
volume is
 \be
\Ls={1\ov\xs^2}.
 \ee

What determines $\xs$?  Obviously it is related to the
correlation length $\xp$ of the order parameter,
specifically of its phase.  But this is not an answer. 
During a second-order phase transition, $\xp$ is
varying rapidly.  Indeed, it is characteristic of
second-order transitions that the equlibrium correlation
length $\xe$ diverges at the critical temperature.  So
we must specify at what time or what temperature $\xs$
should be compared with $\xe$.

An answer to this question has been given by Wojciech
Zurek \cite{Zur85,Zur93,Zur96}, following an earlier
suggestion of mine \cite{Kib80}.  It is clear that in a
real system going through the transition at a finite
rate, the true correlation length $\xp$ can never become
infinite.  In fact, for reasons of causality it can
never increase faster than the speed of light.  So,
beyond the point where $\dot\xe=c$, the adiabatic
approximation, that
$\xp\ap\xe(T)$, ceases to be valid, and instead one may
assume that $\xp$ will be more or less constant until
after the transition, at least to the point where it
again becomes equal to the decreasing $\xe$.  In a
non-relativistic system, it is not the speed of light
that is relevant, but some characteristic speed of the
system.

Zurek has given an alternative argument leading to
essentially the same conclusion, based on a comparison
of the quench rate and relaxation rate of the system.
 
Let us assume that near the transition, the temperature
varies linearly with time, so that
 \be
\ep\equiv 1-{T\ov\Tc}={t\ov\tq}.
 \ee
Here $\tq$ is the quench time.  (We take $t=0$ when
$T=\Tc$.)  The equilibrium correlation length near
$\Tc$ has the form 
 \be
\xe(T)=\xi_0|\ep|^{-\nu},
 \ee
where $\nu$ is a critical index.  In mean field theory,
$\nu=\half$, and this is often an adequate
approximation.  For $^4$He, however, the renormalization
group, gives a more accurate value, $\nu=\case{2}{3}$. 
Similarly the relaxation time $\ta$ diverges at $\Tc$:
 \be
\ta(T)=\ta_0|\ep|^{-\mu},
 \ee
where for $^4$He the critical index $\mu=1$. This is the
phenomenon of {\em critical slowing down}.  The
characteristic velocity is
 \be
c(T)={\xe(T)\ov\ta(T)}={\xi_0\ov\ta_0}|\ep|^{\mu-\nu}.
 \ee
Note that it vanishes at $\Tc$.  In $^4$He, this is the
speed of {\em second sound}, a thermal wave in which the
normal and superfluid components oscillate in antiphase.

Now information about the phase of the order parameter
cannot propagate faster than the speed $c(T)$.  Hence
after the transition the distance over which phase
information can propagate is the sonic horizon
 \be
h(t) = \int_0^t c(T(t'))dt' =
{1\ov1+\mu-\nu}{\xi_0\tq\ov\ta_0}\ep^{1+\mu-\nu}.
 \ee
This becomes equal to the equilibrium correlation length
when
 \be
\ep=\eZ =
\left[(1+\mu-\nu){\ta_0\ov\tq}\right]^{1\ov1+\mu}.
 \ee
The time when this happens is the {\em Zurek time}
 \be
\tZ = \left[(1+\mu-\nu)\ta_0\tq^\mu\right]^{1\ov1+\mu}.
 \ee

It is reasonable to suppose, at least as a first crude
approximation, that {\em at the Zurek time} the
characteristic length scale $\xs$ of the tangle of
strings or vortices should be equal to the correlation
length:
 \be
\xs(\tZ) \sim \xZ=\xe(\tZ) \sim
\xi_0\left(\tq\ov\ta_0\right)^{\nu\ov1+\mu}.
 \ee
Equivalently, we expect the density of strings or
vortices (\ie, the length per unit volume) to be
approximately $1/\xZ^2$, \ie,
 \be
\Ls(\tZ) =
{\ka\ov\xi_0^2}\left(\ta_0\ov\tq\right)^{2\nu\ov1+\mu},
 \label{Lstq}
 \ee
where $\ka$ is a numerical constant of order unity.
Numerical simulations \cite{LagZ97,YatZ98} suggest that
it should in fact be somewhat less than unity, perhaps
of order $0.1$.  Note that in $^4$He, the exponent in
(\ref{Lstq}) is $1\ov2$ in mean field theory, while
using renormalization-group values it is $2\ov3$.  This
is the prediction that has to be tested.

\subsection{Experiments in helium-4}

Zurek \cite{Zur85} initially suggested testing these
predictions in superfluid $^4$He.  Experiments designed
to test his predictions have been performed by Peter
McClintock's group at Lancaster using a rapid pressure
quench.  The experimental sample was contained in a
small chamber that could be rapidly expanded to lower
the pressure, thereby sending it through the lambda
transition into the superfluid phase.  The number of
vortices produced was found by measuring the attenuation
of a second sound signal, generated by a small heater.

The first experiment \cite{Hen+94} did in fact see
evidence of vorticity generated during the quench, at
roughly the predicted level.  However, it was not
conclusive for various reasons.  Vorticity might have
been produced by hydrodynamical effects at the walls. 
Also the capillary tube used to fill the chamber was
closed at the outer end, so that during the expansion
some helium was inevitably injected into the chamber,
possibly again creating vorticity.  Another problem was
that it was not possible to measure the second-sound
attenuation during the first 50 ms after the transition,
so that later readings had to be extrapolated back to
the relevant time.

To overcome these problems, the apparatus was redesigned
to minimize the hydrodynamic effects, and the experiment
repeated \cite{Dod+98}.  Somewhat disappointingly, the
result was null: no vorticity was detected with the
improved apparatus.  One possible explanation for this
is that the vortices produced may simply disappear too
fast to be seen \cite{Riv00,Riv01}.  The rate at which
vorticity dissipates was measured in the rather
different circumstances of vorticity generated by
turbulent flow.  It is not certain that the results can
be carried over to the circumstances of this experiment.

A third version of the experiment, incorporating
further improvements, is now being planned
\cite{Hen+00}.  Results are eagerly awaited.

\subsection{Experiments in helium-3}

There are a number of advantages in using $^3$He rather
than $^4$He.  One is that because the correlation length
is much longer (40 to 100 nm, rather than less than 1
nm), a continuum (Ginzburg--Landau) description
is much more accurate than in $^4$He.  Moreover, the
energy needed to generate a vortex is larger relative to
the thermal energy, so it is easier to avoid extrinsic
vortex formation.  Another advantage is that since the
nuclear spin is non-zero, one can use nuclear magnetic
resonance to count the vortices.

Perhaps the greatest advantage, however, lies in the
fact that one can induce a temperature- rather than
pressure-driven transition.  This is because of another
characteristic of $^3$He, namely that it is a very
efficient neutron absorber, via the reaction
 \be
n + {^3{\rm He}} \to p + {^3{\rm H}} + 764\;{\rm keV}.
 \ee

Two experiments have been done with $^3$He, one in
Grenoble \cite{Bau+96} and one in Helsinki
\cite{Ruu+96}.  Both use $^3$He in the superfluid $B$
phase, and look for evidence of vortices similar to
those in $^4$He.  Both make use of the neutron
absorption reaction, by exposing the helium container to
neutrons from a radioactive source.  Each neutron
absorbed releases 764 keV of energy, initially in the
form of kinetic energy of the proton and triton.  This
serves to heat up a small region to above the transition
temperature.  It then rapidly cools, in a time of the
order of 1 $\mu$s, and goes back through the transition
into the superfluid phase.  During this process we
expect a random tangle of vortices to be generated.

In other respects the experiments are very different. 
The Grenoble experiment \cite{Bau+96}, using a sample of
$^3$He-$B$ at a temperature much less than $\Tc$ was
essentially calorimetry.  The total energy released, in
the form of quasiparticles, following each
neutron-absorption event was measured.  Of the available
764 keV of energy about 50 keV is released in the form
of ultraviolet radiation.  However, the measured energy
was in the range 600 to 650 keV, depending on the
pressure, leaving a considerable shortfall.  This is
interpreted as being the energy lost to vortex
formation.  It is very hard to think of any other
possible interpretation.  

The main feature of the Helsinki experiment
\cite{Ruu+96}, using a sample of $^3$He-$B$ at a
considerably higher temperature, not far below $\Tc$,
was the use of a rotating cryostat.  If a container of
helium is rotated rapidly, vortices are generated at the
walls and migrate to form a central cluster parallel to
the rotation axis.  However, if the rotation is slower,
no vortices can be formed.  In $^3$He-$B$, it is
possible to ensure that no vortices at all are present. 
We then have a remarkable situation.  The normal fluid
component is rotating with the container, but the
superfluid component, which cannot support vorticity, is
completely stationary.  Thus there is a {\em
counterflow} velocity, a difference $\bv=\bvs-\bvn$
between the velocities of the two components.  This
introduces novel hydrodynamic effects; in particular a
superfluid vortex moving relative to the normal fluid
experiences the transverse {\em Magnus force}.  

In consequence vortices above a certain minimum size
$r_0$ and correctly oriented are expanded until they
reach the walls of the container, and then migrate to
join a central cluster parallel to the axis.  The number
of vortices `captured' in this way can be determined by
nuclear magnetic resonance (NMR) measurements.  It is
possible to detect each individual vortex joining the
cluster.

The number of vortices we expect to be captured can be
predicted.  It is essentially the number of vortices
with sizes between the required minimum size $r_0$,
which depends on the counterflow velocity $v$, and the
maximum radius of the bubble.  The size distribution of
loops formed is expected to be scale invariant.  This
leads to a very simple prediction.  There is a critical
velocity $\vcn$ for neutron-induced vortex
formation, which is substantially lower than the critical
velocity $\vc$ for spontaneous vortex formation at the
walls.  If $v>\vcn$, the number of vortices captured
after each neutron-absorption event should have the form
 \be
N = c\left[\left(v\ov\vcn\right)^3-1\right],
 \ee
where $c$ is a calculable constant.  Remarkably enough,
all the dependence on the bulk temperature, the pressure
and the magnetic field is contained in the value of
$\vcn$.  Hence if $N$ is plotted against $v^3$ for
various values of these parameters, one should see a set
of straight lines with a common intercept at $-c$ on the
vertical axis.  This simple prediction does fit the
experimental results very well over a considerable
parameter range, providing good evidence for the
validity of the prediction.

It has also been possible to test the predicted
dependence of $\vcn$ on temperature, namely
$\vcn\propto\ep^{1/3}$.  This again is a good fit to the
data.

\subsection{Experiments in superconductors}

It is particularly interesting to test the predictions
of defect formation in superconductors, because they
provide an example of a {\em gauge} theory.

The first experiments \cite{CarP99} were done by a group
at Technion, using a thin film of the high-temperature
superconductor YBCO.  The film was raised above the
critical temperature by shining a light on it, and then
allowed to cool.  The object of the experiment was to
determine the number of defects formed, in this case
`fluxons' each carrying one quantum of magnetic flux.
What Carmi and Polturak measured, using a SQUID
detector, was actually the {\em net} flux, \ie, the
difference $\De N=N_+-N_-$ between the numbers of fluxons
and antifluxons.  In fact they saw no evidence for any
fluxon formation, with an upper limit of $|\De N|<10$.

This result has to be compared with predictions based on
Zurek's work.  In this case the Zurek length $\xZ$ is
estimated to be about $10^{-7}$ m, so within the 1 cm$^2$
sample we should expect the total number of defects to be
 \be
N=N_++N_-\ap10^{10}.
 \ee
The net flux may be estimated by assuming that the phase
of the order parameter performs a random walk with a step
length of $\xZ$ and a typical angle $\de\sim\pi/2$.  This
suggests that 
 \be
\De N \ap {\de\ov2\pi}\sqrt{L\ov\xZ} =
{1\ov4}\sqrt{L\ov\xZ},
 \ee
where $L\ap 20$ mm is the perimeter of the sample. 
(Note that according to this argument $\De N$ is of order
$N^{1/4}$.)  This yields
 \be
\De N \ap 100,
 \ee
in clear contradiction to the experimental results.  It
should be noted that this prediction is based on
(\ref{Lstq}) with the constant $\ka$ set equal to unity,
so there may be scope for reducing it slightly, though
probably not by enough to remove the discrepancy.

On the other hand, Carmi and Polturak in fact suggest
that the disagreement is {\em more} serious, because in a
gauge theory the mechanism of defect formation is
different and the geodesic rule is unreliable
\cite{RudS93,CopS96,HinR00}, so one should perhaps expect
$\De N$ to be of order $N^{1/2}$, leading to an estimate
$\De N\ap 10^4$ which is obviously in very severe
disagreement with the results.  This is a point that
needs further theoretical study.

However, the same group have also performed another
experiment \cite{Car+00}, with very different results. 
This involved a loop of superconducting wire laid down
in a square-wave pattern across a grain boundary in the
substrate so as to create a series of ${\cal N}=214$
Josephson junctions in series.  As the wire cools it
becomes superconducting before the Josephson junctions
start to conduct, so in effect each segment of wire
between neighbouring junctions is initially a separate
system, so it is reasonable to assume that their phases
are random and uncorrelated.  Hence some flux will be
trapped when the wire eventually becomes a single
superconducting loop.  The experiment revealed an
r.m.s.\ flux of 
 \be
\De N\rms{exp} = 7.4 \pm 0.7.
 \ee
The theoretical prediction in this case would be  
 \be
\De N\rms{th} = {1\ov4}\sqrt{\cal N} = 3.6.
 \ee
It is perhaps rather surprising that the experiment saw
{\em more} flux than predicted.  The authors
suggest that this may again be due to a breakdown of the
geodesic rule with an r.m.s.\ value of $\de$ closer to
$\pi$ than to $\pi/2$.  (Arguably, if $\de$ is uniformly
distributed between $-\pi$ and $\pi$, we should use an
r.m.s.\ value of $\pi/\sqrt3$ rather than $\pi/2$, but
the difference is minimal, leading to
$\De N\rms{th}=4.3$.)  There could also perhaps be a
non-zero phase change along the section of the loop away
from the Josephson junctions.

Recently experiments have been performed by a different
group \cite{Kav+00,Mon+02} on annular Josephson
tunnelling junctions, comprising two rings of
superconducting material separated by a thin layer. 
When the system is cooled through the critical
temperature and the rings become superconducting, one
may expect that the random choice of phase will lead to
trapping of fluxons.  For the experiments done so far
the predicted number trapped is less than one fluxon on
average, which is not ideal.  Nevertheless, they have
detected flux trapping at roughly the predicted level. 
An important feature of this experiment is that it is
possible to vary the quench rate and so test the
dependence of the number of fluxons on the quench rate
$\tq$, as given by (\ref{Lstq}).  The results are
consistent with the Zurek predictions, though the
scatter is large.

\subsection{Discussion}

Experiments with $^3$He, with liquid crystals and with
superconducting loops have all confirmed the basic idea
that defects are formed during rapid phase transitions. 
The best evidence so far that Zurek's predictions of
defect numbers are sound comes from the $^3$He
experiments, though the others are reasonably consistent.

On the other hand, neither the $^4$He experiment nor
that with a superconducting film have shown any evidence
for defect formation. 

At first sight, the discrepancy between the results with
$^4$He and $^3$He may be surprising, but in fact the
differences between the two systems are very great. 
Karra and Rivers \cite{KarR97} have argued that
a very important factor is the great discrepancy between
the widths of the `critical region', below the
critical temperature and above the {\em Ginzburg
temperature} $\TG$ \cite{TilT90a}.  This is the
temperature above which thermal fluctuations are large
enough to create a significant transient population of
thermally excited small vortex loops.  It is given
approximately by the condition that 
 \be
\xe^3(\TG)\De F(\TG) = \kB\TG,
 \ee
where $\De F$ is the difference in free energy between
the `false-vacuum' state with $\ph=0$ and the
broken-symmetry equilibrium state.  Above this
temperature, it appears, the formation of long vortices
is suppressed.  It happens that $^3$He and $^4$He are
very different in regard to the width of the critical
region between $\TG$ and $\Tc$.  In $^3$He it is extremely
narrow; $\TG$ is very close to $\Tc$, at $\ep\ap
10^{-8}$.  In $^4$He, on the other hand, $\TG$ is about
half a degree below $\Tc$.  Karra and Rivers \cite{KarR97}
used thermal field theory, with a Gaussian approximation,
to show that the Zurek predictions should be
approximately valid provided that
 \be
\ep(\TG){\tq\ov\ta_0} \lap 100,
 \label{KRineq}
 \ee
a condition that is very well satisfied for the $^3$He
experiments where the left hand side is about $10^{-5}$
and badly violated for those in $^4$He, where it is
$10^{10}$.

Also puzzling is the discrepancy between the different
experiments in superconductors.  There is some doubt
about how to compute the number of defects formed in a
transition in a theory with a local gauge symmetry. 
There is another mechanism operating in a gauge theory
\cite{HinR00,HinR01}, but if anything this makes the
discrepancy more puzzling because it tends to suggest
that the Zurek prediction of defect numbers would be an
underestimate.  On the other hand, it is worth noting
that the inequality (\ref{KRineq}) is also seriously
violated in the superconducting film experiment, though
whether the argument leading to it is valid in the case
of symmetry breaking in a gauge theory is not clear.

What is clear is that there is as yet no certainty
about when the cosmology-based predictions of defect
numbers are reliable.  Only further experimental and
theoretical work will resolve this question.

\acknowledgements

These lectures were presented at the Summer School on
{\em Patterns of Symmetry Breaking}, supported by NATO
as an Advanced Study Institute and by the European
Science Foundation Programme {\em Cosmology in the
Laboratory}.  I am grateful to several participants for
pointing out errors and suggesting improvements to the
preliminary version.

\end{document}